\documentclass[doublecol]{epl2} 
\usepackage{epsfig,graphics,graphicx,amsmath,amssymb,float}
\usepackage{color,soul}

\title{Confronting $\gamma$-rays from singlet fermionic cold dark matter with the H.E.S.S. data}

\author{R. Moazzemi}

\institute{                    
  Department of Physics, University of Qom, Ghadir Blvd., Qom 37161-46611, I.R. Iran
  
}


\abstract{
	We explore the whole parameter space of the singlet fermionic cold dark matter model with respect to constraints on, first, the relic density and second, gamma-ray lines up to 10 TeV.  We investigate 44000 random sample models which comprehensively scan the parameter space for dark matter mass below 10 TeV, and compare our results with the latest experimental data from H.E.S.S., for the first time. It is showed that, except for the resonance regions,	this indirect detection cannot exclude the parameter space of this model. }

\begin{document}

\maketitle

	\section{Introduction}
It is confirmed that more than 95\% of the universe content is dark; about 70\% dark energy, 25\% dark matter (DM) and only about 5\% is visible (baryonic). During recent decades, several models, beyond the Standard Model (SM) of particle physics, have been proposed to explain this puzzling dark part of the universe.  The most important candidates for cold DM (CDM) are weakly interacting massive particles (WIMPs) which can resolve some   related cosmological and astrophysical problems. The other important candidates are axions and MACHOs. Sterile neutrinos can also be considered as  warm dark matter particles \cite{dodelson}.  Although we have failed to detect DM as a signal in a collider, because of its negligible interactions with SM particles, direct and indirect searches are still possible. The experiments of direct detection, search for observing the recoil energy of atomic nuclei interacting with DM particles passing through the earth. Indirect detection efforts, however, search for possible visible products of decay or annihilation of DM particles. Gamma-rays are one of the final products of which some experiments, such as Fermi-LAT, VERITAS and H.E.S.S., can measure their excess as an indirect detection of DM.
In addition to direct and indirect detection bounds, there is an important bound on the observed relic density of DM. Indeed, the present relic abundance  of CDM is
$\Omega_{\rm{CDM} }h^2=0.120 \pm 0.001$ \cite{Aghanim:2018eyx} where $h\approx0.7$ is the scaled Hubble constant in units
of 100 km/sec/Mpc.

Various scalar WIMPs have been widely  considered in the literature (see for example \cite{zee,sdm1,sdm2,sdm3}). For a review on the subject of WIMP searches and models see for example \cite{arcadi} and for a very recent review on DM through the Higgs portal see \cite{arcadi2}. The situation is the same for  fermionic DM (see for instance \cite{kim1,fdm2,fdm3}) and SUSY DM \cite{susy}. The singlet fermionic cold dark matter (SFCDM) model is also one of an interesting model which consider a simple extension of the SM to a hidden sector where the fermionic DM can interact with the SM sector through a Higgs portal with a typical coupling $g_s$\cite{kim2}. Therefore,  there are two Higgs bosons in this model and one of them  must play the role of the SM Higgs with 125 GeV mass according to 2012 ATLAS and CMS reports \cite{atlas,cms}. Fixing the other Higgs mass to 750 GeV, in ref \cite{ettefaghi3} we study SFCDM parameter space confronting with recent direct detection data for energy values below 1 TeV. As an important conclusion, this model was completely excluded by recent  XENON100 \cite{XENON100}, PandaX II \cite{pandax} and LUX \cite{LUX2016}  data. In anther work \cite{ettefaghi1}, for DM masses below 200 GeV, we analyze SFCDM annihilation to two photons compared with 2010 Fermi-LAT \cite{fermi2010} data, the approximate cross section was used there and some couplings assumed to be fixed. In this work, for the first time, we extend the energies to 10 TeV and use the precise form of cross section and allow all couplings to be free within their valid range. 

On the other hand, the existence of the other Higgs with the mass below 1 TeV has not yet been confirmed by the LHC. Here we study two different sorts of model by fixing the mass of the second Higgs to  1 and 2.5 TeV. We reconsider this model by calculating the cross section of annihilation of DM particles into monochromatic gamma-ray lines. We allow the mass of the singlet fermion be in the range 200 MeV-10 TeV. This mass range is below the unitarity limits which exist in the literature \cite{Griest,smirnov}. The latest experimental data reported by H.E.S.S. are used for constraining the parameter space. 

There are some issues here, to which we should pay more attention. First, this analysis is done for DM particles with masses up to 10 TeV for the first time within this model. Second, for implementing the relic density constraint,  we calculate the leading order of the complete cross section of DM annihilation into SM final products within the perturbation theory, so that, it is important that $g_s$  remains properly less than one. Therefore, if, for a model, the relic density condition gives {$g_s\geq1$} and by this coupling, the model is excluded by direct or indirect detection, we can  be quite confident in excluding the model. In this case, though the perturbation does not work, we are sure that $g_s$ cannot be less than one, hence, {if one carries out the calculation exactly, he certainly will not get a smaller cross section.} On the other hand, for a model with {$g_s\geq1$  whose} cross section is below the experimental data, {this perturbative approach cannot provide reliable  consequences}; we need exact calculations or possibly more accurate {experiments}. Third, in Ref. \cite{adam}   by investigating the available parameter space for diphoton data in the combined LHC run-1 and run-2, authors show that for the second Higgs mass equal to 750 GeV, if it exists, the mixing angle between two Higgs bosons, $\theta$, should be less than 0.01. 
{There exist other constraints in the literature (see for example  \cite{arcadi2,butt,Ilnicka,Carena}) that, of course, are not stronger than what obtained in Ref. \cite{adam}.} Because of such a constraint (though for 750 GeV Higgs), and to have more values below $\theta=0.01$ when scanning  the parameter space randomly, we partition the parameter space into two parts: $\theta<0.01$ and $\theta>0.01$.

It is reasonable that for heavier Higgs this constraint becomes stronger. Therefore, we perform our analysis in two parts: for $\theta<0.01$ and $\theta>0.01$. In addition, we investigate two different masses for the second Higgs; 1 TeV and 2.5 TeV. In each case, to cover the whole parameter space, we investigate 44000 random sample models; 22000 models for  $\theta<0.01$ and 22000 models for  $\theta>0.01$.  We first implement the relic density bound to derive the allowed parameter space, then the gamma-ray bound reported by H.E.S.S. \cite{hess} is used to restrict the available parameter space. 

The letter is organized as follows: after this introduction, we first briefly present the model of SFCDM. Then, we explore the parameter space allowed by the relic density condition. We then, using the annihilation cross section of SFCDM into two photons, perform numerical calculations on thermally-averaged velocity-weighted form of this cross section and compare the results with recent experimental data. Finally, we present our conclusions.

\begin{figure*}
	\includegraphics[width=17cm]{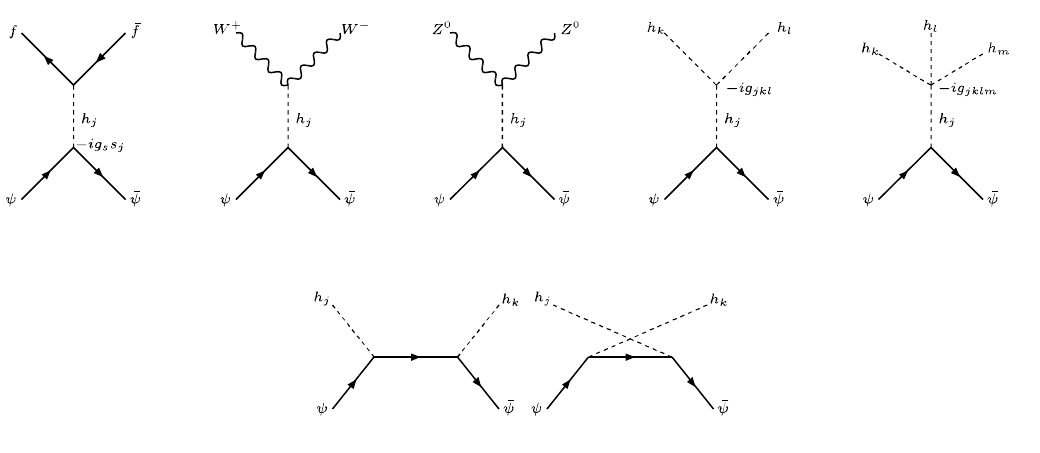}\caption{The Feynman diagrams for the annihilation of  singlet fermion pairs into SM particles, two and three Higgs bosons at tree level. The vertex factor of three (four) Higgs boson lines,  $-ig_{ijk}$ ($-ig_{ijkl}$), is symmetric under permutations of their subscripts. For three Higgs bosons in final state only the dominant Feynman diagrams are shown. Obviously, the first row is due to the $s$-channel while the second row indicates the $t$- and $u$-channels. The vertices have been introduced in Ref. \cite{ettefaghi}}.\label{fig11.}
\end{figure*}

\vspace{2cm}

\section{The model}  \label{model}
We start with the following  renormalizable Lagrangian which has been developed in Ref. \cite{kim1}:
\begin{equation} \label{sfcdm}
{\cal L}_{\rm{SFCDM}}={\cal L}_{\rm{SM}}+{\cal L}_{\rm{hid}}+{\cal L}_{\rm{int}},
\end{equation}
where ${\cal L}_{\rm{SM}}$ is the SM Lagrangian, ${\cal L}_{\rm{hid}}$ denotes  the hidden sector Lagrangian which includes DM and ${\cal L}_{\rm{int}}$ is for interaction between  these two sectors.  With the following definitions, this Lagrangian describe the most minimal renormalizable extension of the SM which includes a singlet fermion field, $\psi$, as CDM:
\begin{eqnarray}\label{lhid}
{\cal L}_{\rm{hid}}&=&{\cal L}_\psi+{\cal L}_{\rm{S}}-{g_s \overline \psi  \psi S},\\
{\cal L}_{\rm{int}}&=&-\lambda_1H^\dag HS-\lambda_2H^\dag
HS^2,
\end{eqnarray}
with
\begin{eqnarray}{\cal L}_\psi&=&\bar{\psi}(i\partial\!\!\!/-m_{\psi_0})\psi,\\
{\cal L}_{\rm{S}}&=&\frac 1 2 (\partial_\mu S)(\partial^\mu S)-\frac{m_0^2}{2}S^2-\frac{\lambda_3}{3!}S^3-\frac{\lambda_4}{4!}S^4.
\end{eqnarray}
where a singlet Higgs $S$, in addition to the usual Higgs doublet $H$, is used as a mediator between SFCDM and the SM particles. As it is obvious from Eq. \eqref{lhid}, the last term violates the usual $Z_2$ symmetry so that we cannot invoke this symmetry. Respecting the relic abundance condition, the singlet fermion should have a very weak interaction with the SM particles. After spontaneous symmetry breaking we have
\begin{equation}
H=\frac{1}{\sqrt{2}}\left(
\begin{array}{c}
0 \\
h+v_0
\end{array}\right),
\end{equation}
and
\begin{equation}
S=s+x_0,
\end{equation}
where $v_0$ and $x_0$ are the vacuum expectation values (VEVs) of the SM Higgs and singlet Higgs, respectively. Therefore, the fields $h$ and $s$ are naturally the fluctuations around the VEVs. Diagonalizing the mass matrix as follows gives the mass eigenstates:
\begin{eqnarray}
h_1=s \sin\theta +h \cos\theta ,\nonumber \\
h_2=s \cos\theta -h \sin\theta .
\end{eqnarray}
The mixing angle $\theta$ can be written in terms of the parameters in the Lagrangian (\ref{sfcdm}). The maximal mixing occurs at  $\theta=\pi/4$, so that we can think of $h_1$ as the SM Higgs-like scalar, and $h_2$ as the singlet-like one. { The mixing affects the SM-like Higgs couplings to both fermions and gauge bosons in an identical fashion (but suppressed by the factor $\sin\theta$) and all SM couplings are suppressed by the factor $\cos\theta$. In addition, $\sin\theta$ has to be small enough to maintain the dominantly doublet nature of $h$.} After symmetry breaking, the mass of the singlet fermion becomes $m_\psi=m_{\psi_0}+ g_sx_0$, which is an independent parameter in the model. The VEV of our singlet Higgs, $x_0$, is completely determined by minimization of the total potential. The SM Higgs mass is fixed to {125.09} GeV  according to the 2012 ATLAS \cite{atlas} and CMS \cite{cms} reports. Therefore, we encounter eight independent parameters, in addition to the SM ones, in this model: singlet fermion mass $m_\psi$, second Higgs mass $m_{h_2}$, coupling constants $g_s, \lambda_1$, $\lambda_2$, $\lambda_3$, $\lambda_4$ and Higgs mixing angle $\theta$ .

\section{ The relic density}‎\label{sec3}
In the early universe when the interaction rate of a particle species drops below the expansion rate of the universe, it gets out of the equilibrium and its  density number in the comoving volume does not change. This is called the `freeze-out' mechanism. A WIMP can be  thermally  produced  through a  `freeze-out' mechanism. A singlet fermion pair, $\overline \psi  \psi $, can annihilate into the SM fermions, the gauge bosons and the Higgs bosons. The Boltzmann equation gives the evolution of the density number $n_\psi$ of a singlet fermion:

\begin{equation}‎
‎\frac{dn_\psi}{dt}+3Hn_\psi=-\left\langle\sigma_{\rm{ann}}v\right\rangle\left[n_\psi ^2-\left(n_\psi ^{\rm{eq}}\right)^2\right],
\end{equation}
where $H$ is the Hubble constant, $\left\langle {\sigma_{\rm{ann}}v } \right\rangle$ is the thermal average of the annihilation cross section times the relative velocity, and $n_\psi ^{\rm{eq}}$ is the equilibrium density number of $\psi$. As the expansion rate of the universe exceeds the interaction rate of our WIMPs, this species falls out of equilibrium. Now its relic density $\Omega_\psi h^2$ which is defined as the ratio of the present and critical densities, is written roughly as follows:
\begin{equation}
\Omega_\psi h^2\approx\frac{(1.07\times10^9)x_F}{\sqrt{g_*}M_{\rm{Pl}}(\rm{GeV})\left\langle\sigma_{\rm{ann}}v\right\rangle},
\end{equation}
where  $x_F=m/T_F$ is the inverse freeze-out temperature determined by the following iterative equation:
\begin{equation}
x_F=\ln\left(\frac{m_\psi}{2\pi^3}\sqrt{\frac{45M_{\rm{Pl}}}{2g_*x_F}}\left\langle\sigma_{\rm{ann}}v\right\rangle\right),
\end{equation}
and $g_*$ denotes the effective degrees of freedom for the relativistic quantities in equilibrium \cite{colb}. The  thermally averaged annihilation cross section times the relative velocity, $\left\langle {\sigma_{\rm{ann} }v} \right\rangle$ is 
\begin{eqnarray}‎
\nonumber‎\left\langle\sigma_{\rm{ann}}v\right\rangle&=&\frac{1}{8m_\psi^4 TK_2^2\left(\frac{m_\psi}{T}\right)}\int_{4m_\psi^2}^\infty \sigma _{\rm{ann}} \left( s \right)\left(s-4m_\psi^2\right)\\&&\hspace{2.5cm}\times\sqrt{s}K_1\left(\frac{\sqrt{s}}{T}\right)ds,\label{tacr}
\end{eqnarray}‎
where $K_{1,2}(x)$ are the modified Bessel functions \cite{gondolo}. The cross section for annihilation of a singlet fermion pair into SM final states and Higgs bosons, $\sigma _{\rm{ann}} $, was reported in Ref. \cite{ettefaghi}. Fig. \ref{fig11.} shows the tree level Feynman diagrams for this cross section. 	
As a general rule, the final state $W^+ W^-$ tends to dominate the total annihilation
cross section whenever such channel is open. A light dark matter mass, $m_\psi<m_W$, will annihilate mainly into the $b\bar{b}$  final state. An intermediate mass singlet, $m_W<m_\psi<m_t$, annihilates mostly into $W^+ W^-$, with additional contributions from $Z^0 Z^0$ and, if allowed, $h_i h_j$. For a heavier singlet, $m_\psi>m_t$, the pattern is similar, as the $t\bar{t}$  channel gives a non-negligible but subdominant contribution \cite{yaguna}.

{To study the parameter space of this model, we first focus on the role of couplings $\lambda_1, \lambda_2, \lambda_3$ and $\lambda_4$. In Ref. \cite{ettefaghi} it is showed that changing in these parameters does not significantly affects cross section. However, to have a more comprehensive analysis, we let their dimensionless form to be freely chosen from the interval [0,0.5], with respect to perturbation issues.}

\begin{figure*}
\hspace{2cm}	$	\begin{array}{c}
	\includegraphics[width=14cm]{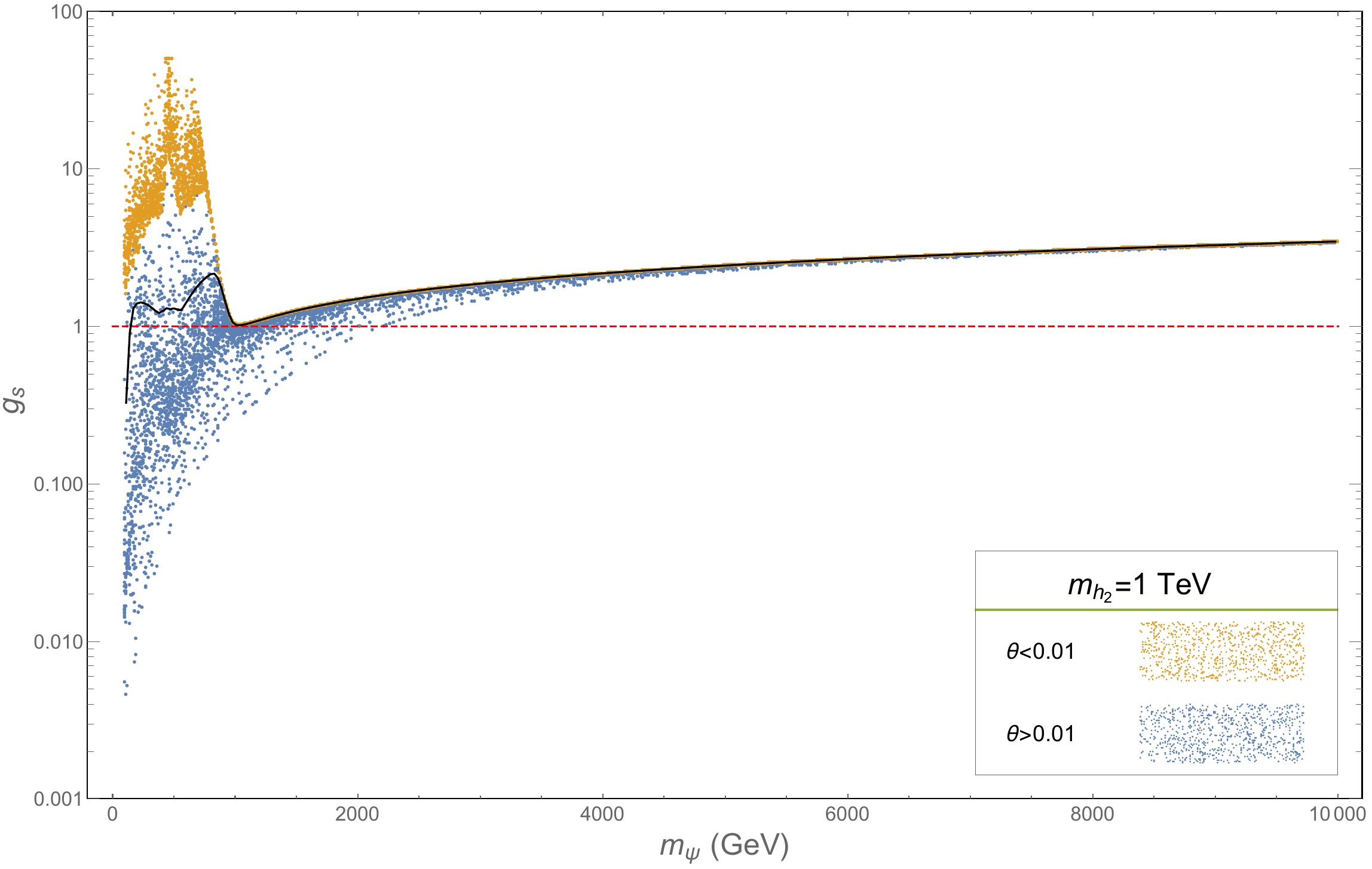}\\
	\includegraphics[width=14cm]{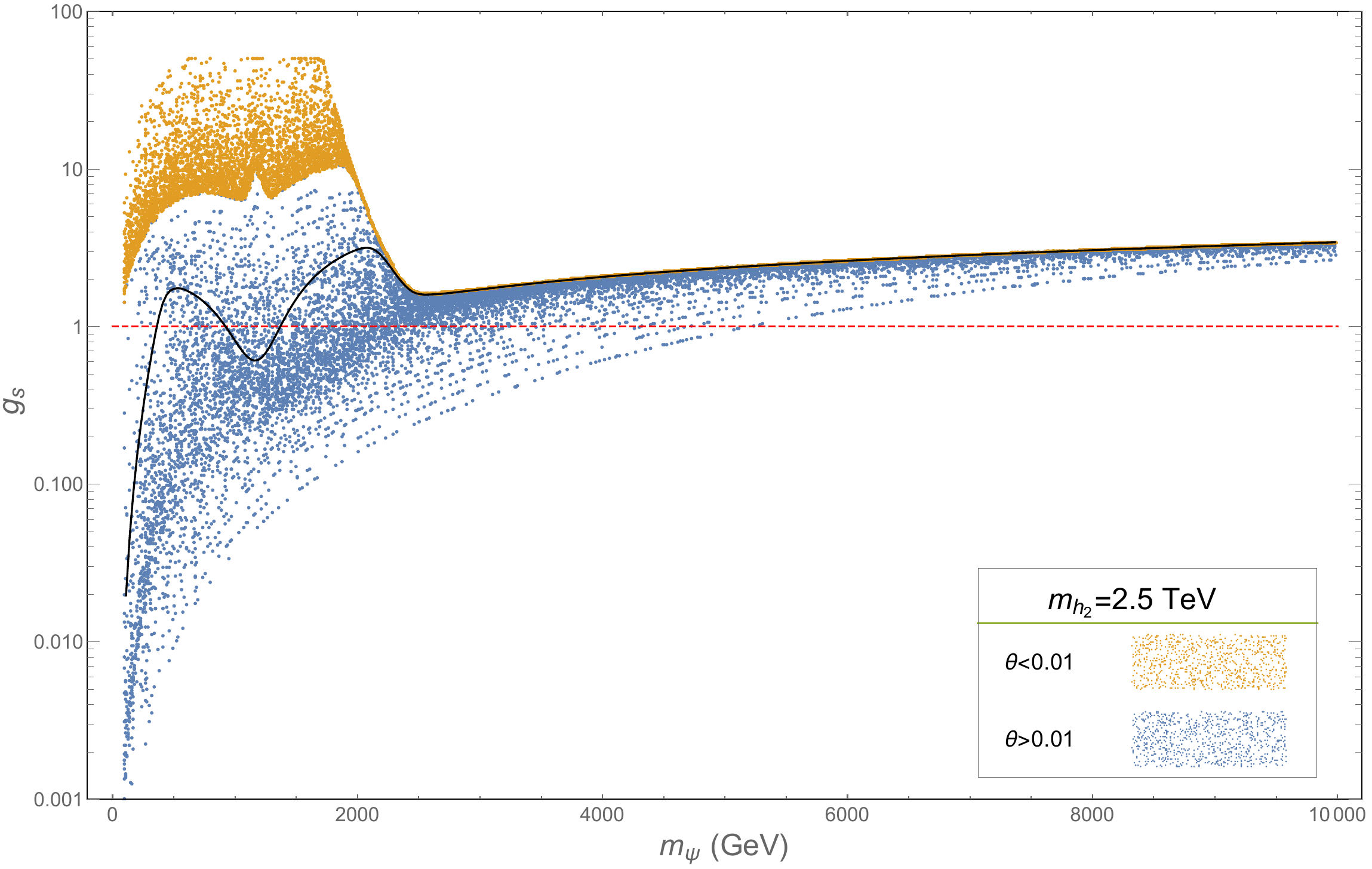}	
	\end{array}$
	\caption{The coupling constant $g_s$ for 44000 random sample models (22000 models with $\theta<0.01$ and 22000 with  $\theta>0.01$) which satisfy the relic density condition with $m_{h_2}=1$ TeV (top) and $m_{h_2}=2.5$ TeV (bottom). The solid line shows the case $\lambda_i=0.1$ and $\theta=0.1$. \label{gs.}}
\end{figure*}

The observed relic abundance  of CDM is  $0.120 \pm 0.001$ \cite{Aghanim:2018eyx}. Above 1 TeV, this limit for dark matter is stronger than  $b\bar{b}$ and $t\bar{t}$ constraints (see e.g. \cite{hbjin}). 
 For two different values of  $m_{h_2}=1$ TeV and 2.5 TeV, we investigate 44000 sample models randomly to cover the parameter space. Each of these two sets includes two parts corresponding to the mixing angle values $\theta<0.01$ and   $\theta>0.01$. Satisfying the relic density constraint, we derive the relative coupling $g_s$. The results are shown in Fig. \ref{gs.}. Since we work at tree level in perturbation theory, all couplings should be less than one. Therefore, the values of $g_s>1$ in Fig.  \ref{gs.} are not consistent with perturbation expansion.

\section{Annihilation to photon pair}\label{sec4}
The annihilation of a pair of SFCDM into a pair of photons, at leading order, occurs by a Higgs particle through the $s$-channel. Although the photon is massless and cannot couple to the Higgs, the $H\gamma\gamma$ vertex can be generated with loops involving massive particles. The leading order Feynman diagrams for this process are shown in Fig. \ref{dm2ph}. The cross section is written as follows:
\begin{figure}[H]
	\includegraphics[width=8cm]{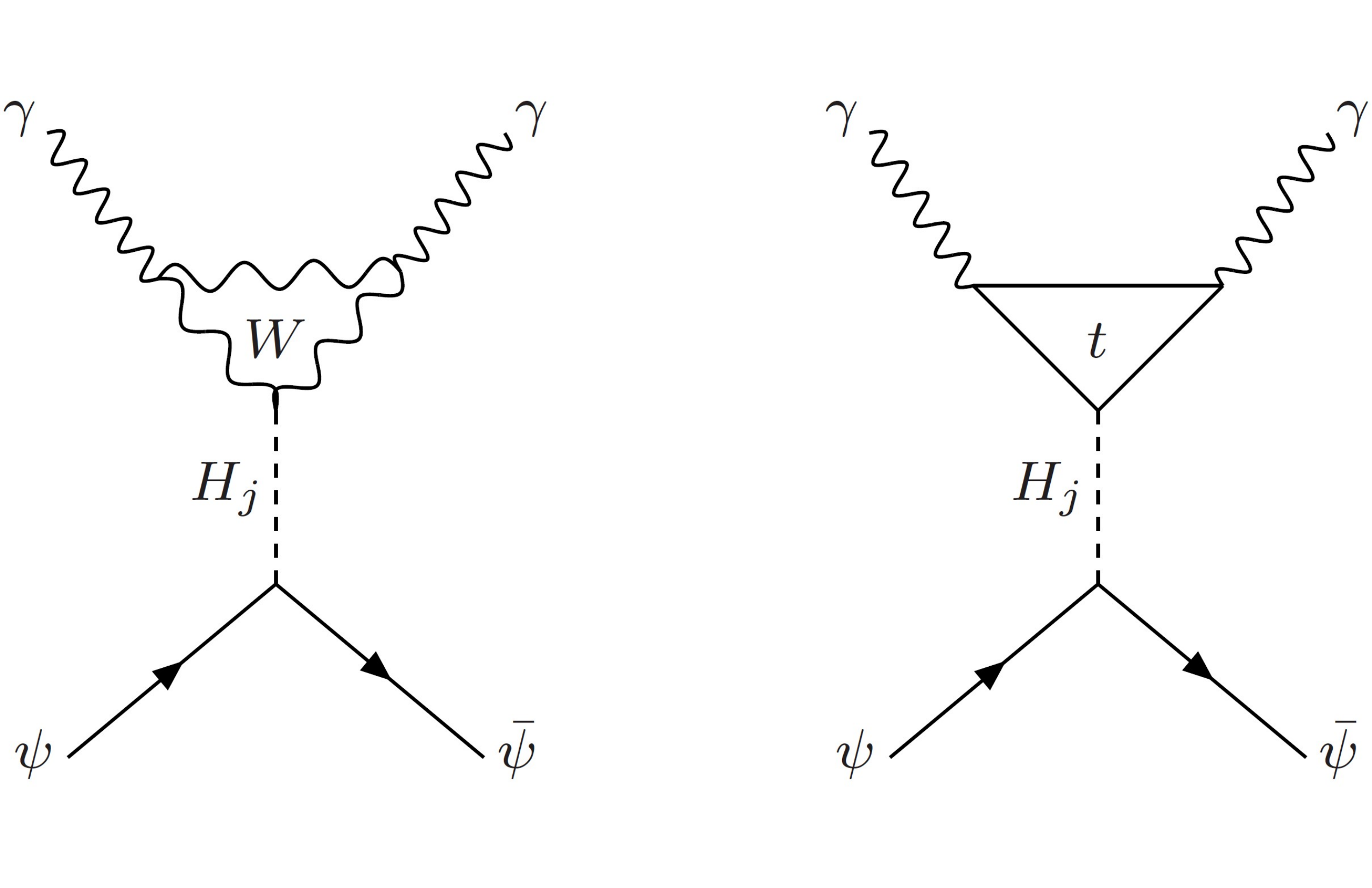}
	\caption{The dominant Feynman diagrams for the annihilation of a  SFCDM pair into monochromatic gamma-ray lines.\label{dm2ph}}
	\label{fig1}
\end{figure}
\begin{equation}\label{2phcs}
\sigma v_{\gamma \gamma} = \frac{1}{{8\pi s}}\frac{1}{4}\sum\limits_{\rm{spins}}
{{{\left| {{M_{\bar \psi \psi  \to \gamma \gamma }}} \right|}^2}} ,
\end{equation}
where $\sqrt{s}$ is the energy in the center of mass frame, and
\begin{equation}\label{2phm}
{M_{\bar \psi \psi  \to \gamma \gamma }} =
\sum\limits_{j = 1,2} {{{\bar v}}(p)i{g_s}{s_j}{u}(p)} \frac{i}{{s - {m^2_{{h_j}}} - i{m_{{h_j}}}
		{\Gamma _j}}}{M_{{h_j} \to \gamma \gamma }},
\end{equation}
where ${s_1}$ and ${s_2}$
denote $\sin \theta $ and $\cos \theta $, respectively,
and ${M_{{h_j} \to \gamma \gamma }}$ the amplitude for the decay of a Higgs into two photons. Here,
${\Gamma _2} = {\Gamma _{{{hhh} }}}+{\Gamma _{{hh} }} + {\Gamma _{{\rm{SM}}}}$ is the sum of three decay rates for the heavier Higgs;
first, the decay rate of a Higgs to three lighter Higgs particles,
$\Gamma_{hhh}=\frac{1}{2}\frac{1}{256\pi^3}g^2_{2111}m_{h_2}	$, second, to two other ones,  $\Gamma_{hh}=\frac{g^2_{211}}{32m_{h_2}\pi}\sqrt{1-\frac{4m_{h_2}^2}{m_{h_2}}}	$ and third, to the SM particles, ${\Gamma _{{\rm{SM}}}}$. For the lighter Higgs only the decay to the SM particles is allowed i.e. ${\Gamma _1} =  {\Gamma _{{\rm{SM}}}}$. We can write ${M_{{h_j}	\to \gamma \gamma }}$ as follows \cite{shifman,gunion}:
\begin{equation}\label{2phh}
{M_{{h_j} \to \gamma \gamma }} = \frac{{\sqrt {1 - s_i^2} \alpha gs}}
{{8\pi {m_W}}}\left[3{\left(\frac{2}{3}\right)^2}{F_t} + {F_W}\right],\quad \mbox{for}\quad i\neq j
\end{equation}
where
${F_t} =  - 2\tau [1 + (1 - \tau )f(\tau )]$, $ {F_W} = 2 + 3\tau  + 3\tau (2 - \tau )f(\tau ).$ and $g$ is the weak coupling constant.
Here, $\tau  = 4{m_i}^2/s$ with $i=t,W$ and
\begin{equation}
f(\tau ) = \left\{ \begin{array}{ll}
{\left( {{{\sin }^{ - 1}}\sqrt {1/\tau } } \right)^2},&\qquad{\rm{for}}\quad\tau  \ge {\rm{1   }}\\
- \frac{1}{4}\left( {\left. {{\rm{ln}}\frac{{1 + \sqrt {1 - \tau } }}
		{{1 - \sqrt {1 - \tau } }} - i\pi } \right)^2} \right.&\qquad{\rm{for}}\quad\tau {\rm{ < 1 }.}
\end{array} \right.
\end{equation}

\begin{figure*}
\hspace{2cm}	$	\begin{array}{c}
	\includegraphics[width=14cm]{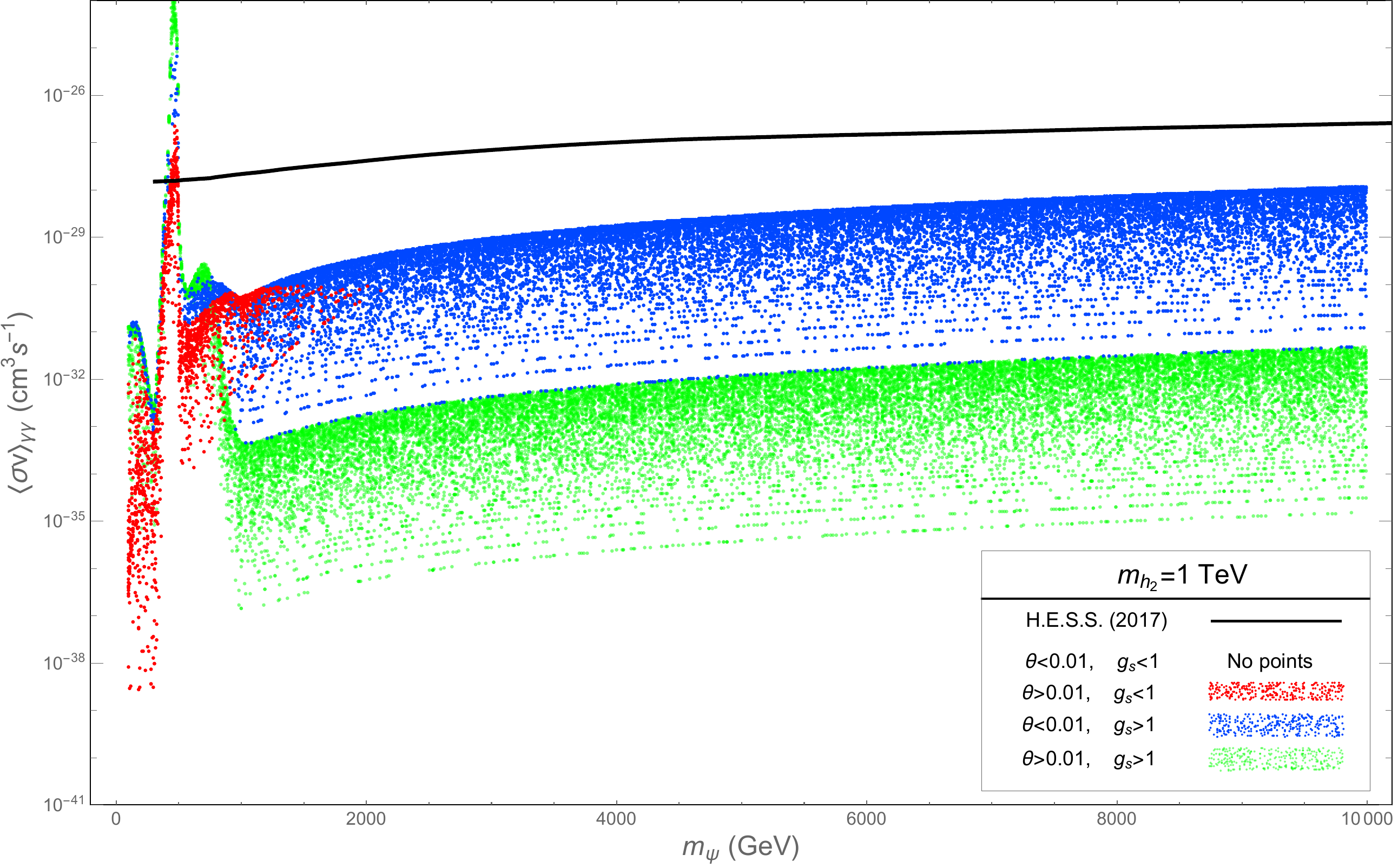}\\
	\includegraphics[width=14cm]{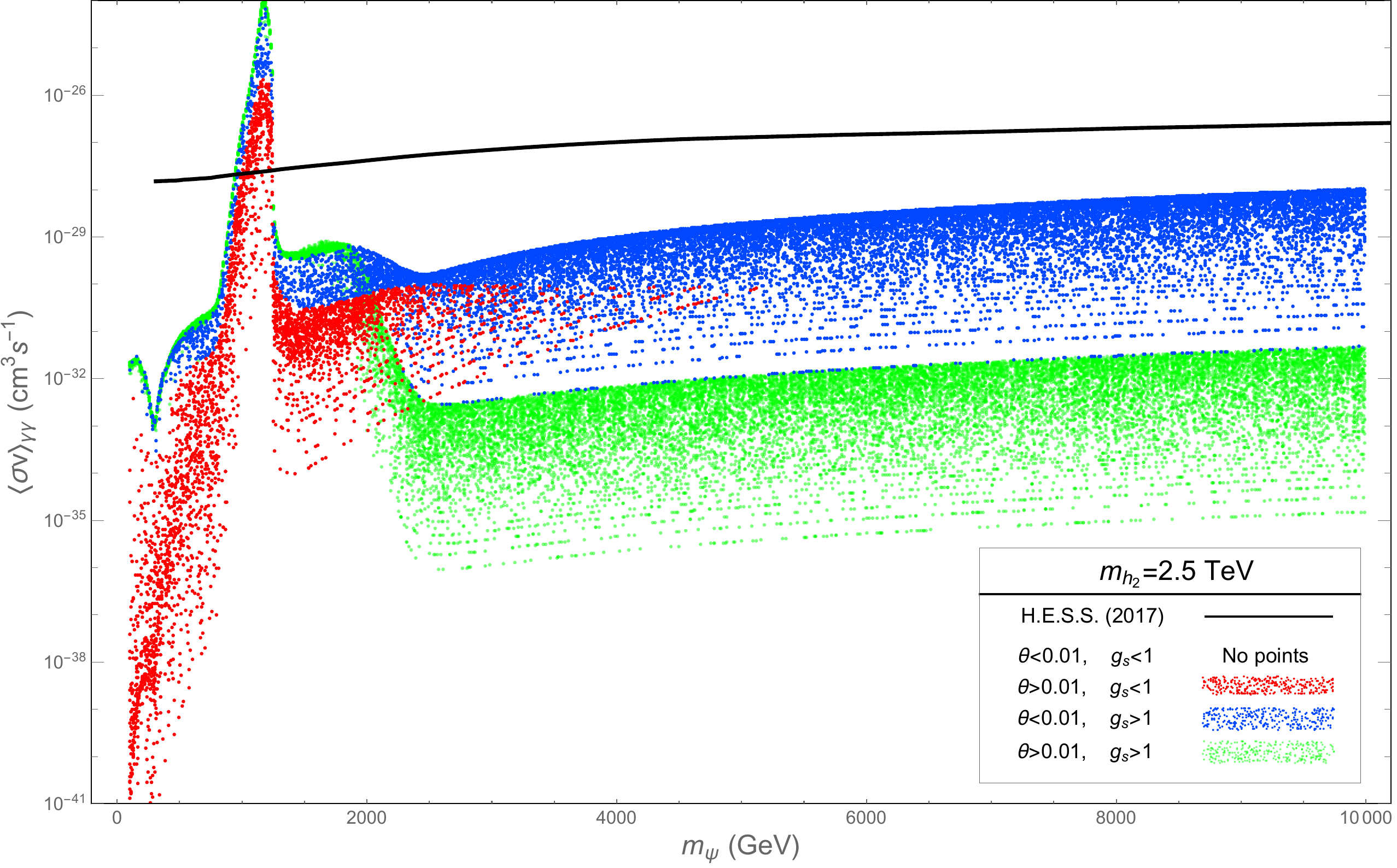}	
	\end{array}$
	\caption{ Thermally averaged
		velocity-weighted annihilation cross section of a SFCDM pair to two photons in terms of the DM mass for $m_{h_2}=1$ TeV (top) and $m_{h_2}=2.5$ TeV (bottom). Solid line shows the 2017 H.E.S.S. upper limit for the Einasto profile \cite{hess}.\label{cs}}
\end{figure*}

We can then derive the  thermally-averaged
velocity-weighted annihilation cross section to  diphoton $\langle \sigma v \rangle_{\gamma \gamma} $, by putting the Eq. (\ref{2phcs}) in Eq. (\ref{tacr}). For all 44000 models, which we considered in the previous section for consistency with relic density, we derive $\langle \sigma v \rangle_{\gamma \gamma} $ and illustrate them in Fig. \ref{cs}. This figure also shows the latest H.E.S.S. \cite{hess} bounds for the Einasto profile. From this figure we see that, there are no points in the parameter space with $\theta<0.01$ which satisfy  the relic density condition along with perturbation criteria. Moreover,
the resonance regions inevitably are excluded.  
{Note that, as we explained in the Introduction, if some region is excluded even with coupling greater than one, we can be sure about this exclusion.}  For $\theta>0.01$ there are also some regions of the parameter space close to the resonance with cross sections near to the upper limits that may be excluded with future more precise experiments. {Such models may be also excluded by enforcing the upper bound on the mixing angle $\theta$.}
{On the other hand, where we are not very close to the resonance region, models with $\theta>0.01$ and $g_s\geq1$ which still seem consistent with current experimental data, require a more comprehensive analysis in order to conclude whether or not these SFCDM models are still allowed.}
 For $m_\psi\gtrsim2m_{h_2}$ there are no regions with $g_s<1$, therefore, the perturbation theory used for this analysis is not suitable and maybe the effective theories here can give us more detail.

\section{Summary and conclusions}
In this letter we have studied the parameter space of the SFCDM model, up to 10 TeV, by computing the leading order of the thermally averaged  annihilation cross section  times relative velocity, within the perturbation theory, and, by calculating the annihilation cross section of a DM pair {into} two gamma-ray lines, we have also  confronted it with the latest  H.E.S.S. data. We have used the most minimal and renormalizable extension of the SM for the SFCDM model. In this model, one adds a singlet fermion as CDM and a new scalar Higgs as a mediator to the SM content. We have scanned the relevant parameter space with 88000 sample models; 44000 for $m_{h_2}=1$ TeV and 44000  for  $m_{h_2}=2.5$ TeV. In each case, we have separated  the parameter space into two parts corresponding to $\theta<0.01$ and $\theta>0.01$. We have  then imposed the relic density condition and found the corresponding $g_s$ and illustrated it in Fig. \ref{gs.} .  The thermally averaged
velocity-weighted annihilation cross section is represented in Fig. \ref{cs}.  As a result we see that,  for $\theta<0.01$ there are no regions where the perturbation theory used here works properly. Such models must be considered effectively. This becomes important if the results of Ref. \cite{adam}, that put the upper limit 0.01 for the mixing angle, also hold for $m_{h_2}>750$ GeV. From Fig. \ref{cs} we see that, except for resonance regions, the cross sections fall below the bound and hence the recent indirect detections could not exclude the model{, though models with $g_s\geq1$ required more comprehensive analysis to conclude whether or not these are still allowed.}


\acknowledgments
The author would like to sincerely thank M.M. Ettefaghi  for very fruitful and constructive discussions.

\end{document}